\documentclass[twocolumn,usenatbib,amsmath,amssymb,amsbsy]{mn2e}

\usepackage[T1]{fontenc}
\usepackage{aecompl}

\usepackage{natbib}
\usepackage{amsbsy}
\usepackage{amssymb}
\usepackage{amsmath}

\usepackage{graphicx}

\begin{document}

\title[Evolution of  magnetar  oscillations] {Three evolutionary paths for magnetar oscillations}

\author[Glampedakis \& Jones]{K. Glampedakis$^{1,2}$ \& D. I. Jones$^3$ \\
  \\
  $^1$ Departamento de F\'isica, Universidad de Murcia, Murcia, E-30100, Spain \\
  $^2$ Theoretical Astrophysics, University of T\"{u}bingen, Auf der Morgenstelle 10, T\"{u}bingen, D-72076, Germany \\
  $^3$ Mathematical Sciences and STAG Research Centre, University of Southampton, Southampton SO17 1BJ, UK}

\maketitle

\begin{abstract}

Quasi-periodic oscillations have been seen in the light curves following several magnetar giant flares.  These oscillations are of great interest as they probably provide our first ever view of the normal modes of oscillation of neutron stars.  The state-of-the-art lies in the study of the oscillations of elastic-magnetic stellar models, mainly with a view to relating the observed frequencies to the structure and composition of the star itself.  We advance this programme by considering several new physical mechanisms that are likely to be important for magnetar oscillations.  These relate to the superfluid/superconducting nature of the stellar interior, and the damping of the modes, both through internal dissipation mechanisms and the launching of waves into the magnetosphere.  We make simple order-of-magnitude estimates to show that both the frequencies and the damping time of magnetar  oscillations can evolve in time, identifying three distinct `pathways' that can be followed, depending upon the initial magnitude of the mode excitation.  These results are interesting as they show that the information buried in magnetar QPOs may be even richer than previously thought, and motivate more careful examination of magnetar light curves, to search for signatures of the different types of evolution that we have identified.

 \end{abstract}


\newcommand{\be}{\begin{equation}}
\newcommand{\ee}{\end{equation}}
\newcommand{\bear}{\begin{eqnarray}}
\newcommand{\eear}{\end{eqnarray}}
\newcommand{\n}{\hat{n}}
\newcommand{\xp}{x_{\rm p}}
\newcommand{\lam}{\hat{\lambda}}
\newcommand{\rx}{{\rm x}}
\newcommand{\ry}{{\rm y}}
\newcommand{\rn}{{\rm n}}
\newcommand{\rp}{{\rm p}}
\newcommand{\re}{{\rm e}}
\newcommand{\rc}{{\rm c}}
\newcommand{\rv}{{\rm v}}
\newcommand{\rnp}{{\rm np}}
\newcommand{\rpn}{{\rm pn}}
\newcommand{\rnn}{{\rm nn}}
\newcommand{\rpe}{{\rm pe}}
\newcommand{\rpp}{{\rm pp}}
\newcommand{\mut}{\tilde{\mu}}
\newcommand{\dmut}{{\delta \mut}}
\newcommand{\dv}{{\delta v}}
\newcommand{\dw}{{\delta w}}
\newcommand{\bw}{\mathbf{w}}
\newcommand{\bv}{\mathbf{v}}
\newcommand{\bF}{\mathbf{F}}
\newcommand{\bB}{\mathbf{B}}
\newcommand{\bb}{\mathbf{b}}
\newcommand{\bu}{\mathbf{u}}
\newcommand{\en}{\varepsilon_\rn}
\newcommand{\ep}{\varepsilon_\rp}
\newcommand{\eb}{\bar{\varepsilon}}
\newcommand{\ex}{{\varepsilon_\rx}}
\newcommand{\ey}{{\varepsilon_\ry}}
\newcommand{\epstar}{\varepsilon_\star}
\newcommand{\cW}{{\cal W}}
\newcommand{\cL}{{\cal L}}
\newcommand{\cE}{{\cal E}}
\newcommand{\rL}{{\rm L}}
\newcommand{\rM}{{\rm M}}
\newcommand{\rP}{{\rm P}}
\newcommand{\rA}{{\rm A}}
\newcommand{\rT}{{\rm t}}
\newcommand{\cT}{{\cal T}}
\newcommand{\cI}{{\cal I}}
\newcommand{\cR}{{\cal R}}
\newcommand{\cA}{{\cal A}}
\newcommand{\cH}{{\cal H}}
\newcommand{\cJ}{{\cal J}}
\newcommand{\cF}{{\cal F}}
\newcommand{\cC}{{\cal C}}
\newcommand{\cN}{{\cal N}}
\newcommand{\cB}{{\cal B}}


\section{Introduction}
\label{sec:intro}

Amongst the various known incarnations of neutron stars, a magnetar is undoubtedly one of the most extreme.
These objects are believed to be endowed with the strongest magnetic fields in Nature, somewhere in the range of $10^{14}$--$10^{15}\,\mbox{G}$
\citep{DT,TD95}; this property alone is responsible for the different astrophysical signature of magnetars that sets them apart from other neutron stars. 
In observational parlance, magnetars manifest themselves as Soft Gamma-ray Repeaters (SGRs) and Anomalous X-ray Pulsars (AXPs), with high energy emission in the X and Gamma ray wavelengths (see \citet{woods} for a review). 

Of the two classes, SGRs are the more prolific objects in terms of high-energy outbursts. The bursts come in
many shapes and forms, but the most spectacular events are the so-called \emph{giant flares}. These release an enormous 
amount of energy ($\sim 10^{44-46}\,\mbox{ergs}$), several orders of magnitude higher than the quiescence SGR luminosity. 
The only energy reservoir capable of fueling such costly energy output is the ultra-strong magnetic field~\citep{TD95,TD01}. 

Only three such giant flares have been observed to date, all in different objects:  SGR 0526-66 (March 1979), SGR 1900+14 (August 1998)
and SGR 1806-20 (December 2004).
A major discovery that came with these events is the detection of quasi-periodic oscillations (QPOs) in the signal~\citep{barat,israel05,SW05,SW06,WS06,HNK11}. 
Since then, these observations have triggered a frenzy among neutron star theorists because they are taken as strong evidence for excited 
neutron stars oscillations -- indeed, the first evidence of this kind (an alternative interpretation of the QPOs could be based on magnetospheric
physics, but this has attracted much less attention so far).

The early days of  the ``magnetar asteroseismology"  project were an attempt to explain the observed QPOs as elastic (seismic)
modes excited in the neutron star crust \citep{duncan98,messios01,piro05,SA07}. This scenario was a very attractive one because (i) giant flares 
are envisaged as the manifestation of a magnetic field instability that in the process fractures the crust and (ii) the low frequency spectrum ($\sim$ tens of Hz)  
of the axial-parity crustal modes is compatible with the presence of several $\lesssim 100\,\mbox{Hz}$ QPOs.  However, it was soon
realized that the real system is likely to display a much more complex behaviour because of the strong magnetic coupling between the crust and the
core~\citep{levin06,toypaper}.

In effect, this coupling leads to global magnetic oscillations in which both the core and the crust may participate. To make things worse,
the type of oscillations most relevant to magnetar QPOs (axisymmetric and axial Alfv\'en modes) could exist in the form of a \emph{continuum} 
of frequencies, rather than a set of discrete frequencies~\citep{levin07,vhoven08, sotani08, cerda09, colaiuda09, colaiuda11,gabler11,gabler12}. 

This Alfv\'en continuum is now known to be very efficient in redistributing the energy of an initially excited crust mode via resonant absorption~\citep{levin07,vhoven08, gabler11,gabler12}. On the other hand, for oscillations of different symmetry (non-axisymmetric, polar, coupled axial-polar modes), 
the frequency spectrum may be discrete, e.g. \citet{sotani08,lander11,colaiuda12}.

The participation of the stellar core in the models of QPOs has highlighted the need for a better understanding of more exotic
properties, such as  the presence of superfluid and superconducting matter. These have a important impact on the mode spectrum and
should be part of any ``realistic" magnetar model~\citep{SFmagnetar,gabler13,passamonti13}. 

In the recent years significant progress has been achieved with models that include crust-core coupling, superfluidity, and 
axial-polar mode couplings (see references above). Although these models can in principle 
`explain' many of the QPOs (albeit with some fine-tuning) the truth is that the modern version of magnetar asteroseismology is plagued 
with a high degree of degeneracy because of the many degrees of freedom associated with the magnetic field and the superfluid state of matter.  

At the same time, all the effort put in this problem so far has to do with making predictions about the mode frequencies. 
However, the actual QPOs offer more potential ``observables" than just the frequencies: for instance, their individual amplitudes and
life spans will reflect properties of the flare and the stellar oscillations themselves.

The present paper adds one more level of complexity to the problem by considering two important physical mechanisms related to the superfluid/superconducting nature of the stellar interior, which have received little attention thus far.  Both are made possible by the rather large amplitude of the oscillations that are expected to be excited by the magnetar flare events.  One concerns the potential destruction of superfluidity when the oscillation amplitude is sufficiently large, an effect considered recently by \citet{GK13}.  This effect will influence both oscillation decay timescales  and (by determining the fraction of the stellar inertia that participates in the oscillation) the frequencies.  The other concerns the interaction between the rotation vortices in the neutron fluid and the magnetic fluxtubes in the proton superconducting fluid.  As has been studied recently~\citep{HGA13}, a sufficiently large oscillation can cause one to cut through the other, adding a strong source of dissipation to the oscillation.  

Depending upon the initial amplitude of the excitation, we will argue that an excited mode could evolve by following three distinct ``pathways", characterized by different frequency evolutions and damping
timescales.    Based on these evolutionary scenarios, we will argue that extra care must be exercised when one attempts to theoretically identify the observed QPO frequencies  and do ``asteroseismology" (i.e. infer the stellar equation of state, the magnetic field intensity and geometry etcetera).  We also demonstrate the potential importance of the launching of magnetic waves into the magnetosphere as a possible damping mechanism for these oscillations.

\textit{Note for the `fast-track' reader:} the paper is structured in a way that allows the independent study of the ``physics" Sections  (results and conclusions 
-- Sections \ref{sec:evolution} \& \ref{sec:conclusions}) without the need of being exposed to any of the technical details on magnetar oscillations
(these are discussed in Sections~\ref{sec:dynamics} \& \ref{sec:damping}).

\section{Evolutionary scenarios for the QPO\small{s}}
\label{sec:evolution}

\subsection{Key magnetic field thresholds} 
\label{sec:key_thresholds}

A prerequisite for formulating ``evolutionary paths'' for magnetar oscillations excited during a flare event is to understand how the induced magnetic field perturbation $\delta B$ measures up against two key magnetic field thresholds.  One threshold concerns the breaking of superfluidity, while the other concerns the breaking of the pinning between the magnetic fluxtubes and the neutron vortices.  We will now estimate the magnetic field perturbation induced by a flare, as well as these two thresholds.  Note that some technical details are drawn from the following Sections.

A magnetar flare is envisaged as a violent event triggered by the cracking of the neutron star crust as a result of magnetic field activity~\citep{TD95,TD01}.  This will excite the star's normal modes, which  can be idealised as falling into two classes, \emph{crustal} modes, confined mainly to the crust, and \emph{Alfv\'en} modes, of a more global character, with predominantly, elastic and magnetic restoring forces, respectively.

According to this model, prior to the fracture the evolving magnetic field exerts a growing strain on the crust; the crust responds elastically
and the resulting quasi-static balance between the magnetic and elastic forces, $F_{\rm mag}$ and $F_{\rm el}$ respectively, 
takes the form (see Section~\ref{sec:dynamics} for more details),
\be
F_{\rm mag} \approx F_{\rm el} \quad \Rightarrow \quad \frac{B \delta B}{4\pi} \approx \mu \psi ,
\ee
where $\psi$ is the dimensionless strain and $\mu$ is the crust's shear modulus. Instead of using the shear modulus itself (which
is a rapidly varying parameter across the crust) it is much more convenient to work with the shear speed
\be
v^2_s = \frac{\mu}{\rho} ,
\label{vs}
\ee   
where $\rho$ is the total density. This quantity has a clear physical meaning (it represents the speed of elastic waves) and 
remains almost uniform throughout the crust (see for instance Fig.~1 in~\citet{GA06}). 

At the moment of the fracture $\psi$ attains the value $\psi_{\rm br}$ and we can obtain a corresponding $\delta B_{\rm br}$:
\be
\label{dBbr}
\frac{\delta B_{\rm br}}{B} \approx \psi_{\rm br}\, \frac{v^2_s}{v^2_\rA} \quad \Rightarrow \quad
\frac{\delta B_{\rm br}}{B} \approx \left (\frac{\psi_{\rm br}}{0.1}\right )\, \frac{\rho_{14} v_{{\rm s},8}^2}{ B^{2}_{15}} ,
\ee
where $v_\rA$ is the Alfv\'en speed for waves in normal (i.e. non-superconducting matter), defined by
\be
v^2_\rA    = \frac{B^2}{4\pi \rho} .
\label{vA1}
\ee
and $v_{{\rm s},8} = v_{\rm s}/10^8\,\mbox{cm}\,\mbox{s}^{-1} $.  To produce the second equation we have used the scalings $B_{15} = B/10^{15}\,\mbox{G}$, $\rho_{14} = \rho/10^{14}\,\mbox{gr}\,\mbox{cm}^{-3}$. In addition, the  breaking strain $\psi_{\rm br}$ has been normalised to the value suggested by the state-of-the-art simulations of~\citet{horowitz09}.   This sudden removal of strain will lead to an unbalanced magnetic stress, so we can expect the excitation of magnetic field oscillations at a level $B(0)$ at around this level:
\be
\delta B(0) \lesssim \delta B_{\rm br} .
\ee
The perturbation would then decay in time, according to the various mechanisms considered later in this paper.

The time evolving magnetic perturbation $\delta B(t)$ is to be compared with the  threshold, $\delta B_{\rm pin}$, which makes contact with the likely presence of superfluid and 
superconducting matter in the outer neutron star core. In that region of the star neutron vortices are expected to be pinned on to the much denser array of proton 
fluxtubes~\citep{sauls}. The magnetars, being slowly spinning and highly magnetised objects, are favoured for harbouring long-term pinned vortices~\citep{GA11}.  
A generic core oscillation excited during a flare will be associated with a non-vanishing velocity difference $\bw$ between the neutron and proton
fluids (see Section~\ref{sec:dynamics} for details). As a result of this velocity difference there will be a Magnus force acting on a unit length segment of 
any individual pinned vortex, doing work against the pinning force exerted on the same segment by the fluxtubes. Pinning cannot be sustained
once $w$ exceeds a critical lag $w_{\rm pin}$ (see \citet{link03} and Section~\ref{sec:cutting} for the detailed form of this quantity). 

For a global magnetar mode the magnetic field provides the main restoring force; in that case we can use eqn.~(\ref{wA1}) (see Section~\ref{sec:dynamics}) 
which is a simple approximate relation between $w$ and $\delta B$ for global Alfv\'en-type oscillations.
We can then obtain the threshold $\delta B_{\rm pin}$ required for vortex unpinning: 
\be
w =  w_{\rm pin} \quad  \Rightarrow \quad 
\frac{\delta B_{\rm pin}}{B} \sim \frac{w_{\rm pin}}{v_\rA} \approx 4 \times 10^{-3} ,
\label{dBpin}
\ee
where now $v_\rA$ is the Alfv\'en speed appropriate for a wave propagating in the superconducting proton fluid (of density $\rho_\rp=x_\rp \rho$ where
$x_\rp$ is the proton fraction):
\be
v_\rA^2 = \frac{H_\rc B}{4\pi \rho_\rp}
\label{vA2}
\ee
and we have used typical values $\rho_{14} =1$, $x_\rp = 0.05$, $H_\rc = 10^{15}\,\mbox{G}$ for the density,
the proton fraction and the (lower) critical superconductivity field, respectively.

The second magnetic field threshold, $\delta B_{\rm SF}$, is related to the possibility of locally suppressing 
neutron superfluidity. This can be achieved once the relative motion between the superfluid and the normal component
exceeds the so-called Landau critical speed. The destruction of superfluidity has been only recently discussed in the context of neutron stars 
by~\citet{GK13} and our own analysis here makes use of their results.

As is described in detail in Section \ref{sec:multifluids},  the relation between $w$ and $\delta B$ is different for oscillations in the crust and in 
the core (see eqns~(\ref{wA1}) and (\ref{wcrust2}) below) and so we  obtain \emph{two} such thresholds. The detailed derivation
is carried out in Section~\ref{sec:destruction}. For global Alfv\'en modes in the core we obtain
\be
\label{dBsf_alfven}
\frac{\delta B_{\rm SF}}{B} \sim 0.08\, w_{{\rm SF}, 7} B_{15}^{-1/2} ,
\ee
where $w_{{\rm SF}, 7} = w_{\rm SF}/10^7\,\mbox{cm}\,\mbox{s}^{-1}$ is the normalized critical speed for the destruction of superfluidity. 
For elastic modes confined in the crust we similarly have
\be
\label{dBsf_crust}
\frac{\delta B_{\rm SF}}{B} \sim \left(\frac{m_\rn^*/m_\rn}{15}\right) \frac{w_{{\rm SF}, 7}}{v_{{\rm s}, 8}} ,
\ee
where  $m^*_\rn$ is the effective mass of the entrained
superfluid neutrons in the crust.

We observe that according to the above formulae we should expect to have
\be
\delta B_{\rm pin} < \delta B_{\rm SF} .
\ee
As we are about to see in the discussion below, the thresholds $\delta B_{\rm pin}$ and $\delta B_{\rm SF} $ 
for the perturbed magnetic field play a key role in understanding the time evolution of magnetar oscillations 
and making predictions about their observational signature. 

That such large perturbations as these might be excited by a flare follows naturally from the Thompson \& Duncan magnetar model, coupled with the very large breaking strains for (certain parts of) the neutron star crust computed using large molecular dynamics simulations \citep{horowitz09}.  The sudden release of a breaking strain $\psi_{\rm br} \sim 0.1$ will induce non-radial crustal motion on length scales of the order of $\psi_{\rm br} R \sim 1$ km.  From the magnetic induction equation (see equation (\ref{induction}) below), this would correspond to a magnetic field perturbation $\delta B / B \sim 0.1$, of the high level required to break the superfluidity, as given by equation (\ref{dBsf_alfven}) above.  The corresponding crustal displacement for breaking the pinning is about two orders of magnitude smaller (see  equation (\ref{dBpin}) above), with amplitude$\sim 10$ m.  Observations themselves show very high levels of fractional modulation of the electromagnetic output, at around the $20\%$ level.  As noted by \citet{dw12}, it is in fact difficult to account for such high levels of modulation, even when beaming of the emission is invoked.  A large mode amplitude tends to reduce this problem, and is certainly consistent with (although not necessarily implied by) the  observations.

\subsection{The three evolutionary pathways}
\label{sec:paths}

\begin{figure}
\begin{center}
\begin{picture}(100,140)(0,-5)

\thicklines

\put(5,42){\bf (C)}
\put(0,-5){\line(1,0){48}}
\put(48,-5){\line(0,1){45}}
\put(48,40){\line(-1,0){48}}
\put(0,40){\line(0,-1){45}}

\put(1,35){$\delta B_{\rm pin} < \delta B(t) < \delta B_{\rm SF}$}
\put(1,28){Neutrons superfluid}
\put(1,21){Cutting occurs $\Rightarrow$}
\put(1,17){strong mutual friction}
\put(1, 10){No mass loading $\Rightarrow$ Alfv\'en}
\put(1, 6){frequency high}
\put(1,0){`Medium' magnetospheric}
\put(1,-4){damping}

\put(70,42){\bf (D)}
\put(65,-5){\line(1,0){41}}
\put(106,-5){\line(0,1){45}}
\put(106,40){\line(-1,0){41}}
\put(65,40){\line(0,-1){45}}

\put(66,35){$\delta B(t) < \delta B_{\rm pin} < \delta B_{\rm SF}$}
\put(66,28){Neutrons superfluid}
\put(66,21){Vortices pinned $\Rightarrow$}
\put(66,17){no mutual friction}
\put(66, 10){No mass loading $\Rightarrow$}
\put(66,6){Alfv\'en frequency high}
\put(66,0){`Weak' magnetospheric}
\put(66,-4){damping}

\put(2,82){\bf (B)}
\put(10,60){\line(1,0){43}}
\put(53,60){\line(0,1){47}}
\put(53,107){\line(-1,0){43}}
\put(10,107){\line(0,-1){47}}


\put(11,103){$\delta B_{\rm pin} < \delta B_{\rm SF} < \delta B(t)$}
\put(11,96){Superfluid (partially)}
\put(11,92){destroyed $\Rightarrow$  `weak'}
\put(11,88){mutual friction}

\put(11, 82){Mass loading $\Rightarrow$ Alfv\'en}
\put(11, 78){frequency  low}

\put(11,72){Mode coupling?}
\put(11,66){`Strong' magnetospheric}
\put(11,62){damping}

\put(32,127){\bf (A)}
\put(40,117){\line(1,0){40}}
\put(80,117){\line(0,1){20}}
\put(80,137){\line(-1,0){40}}
\put(40,137){\line(0,-1){20}}
\put(50,129){{\bf FLARE!}}
\put(48,122){$\delta B(0) \lesssim  \delta B_{\rm br}$}

\put(45,117){\line(-1,-1){10}}
\put(45,117){\vector(-1,-1){7}}
\put(35, 112){\bf 1}

\put(25,60){\line(0,-1){20}}
\put(25,60){\vector(0,-1){12}}
\put(28,50){$t \sim t_{\rm SF}$}

\put(60,117){\line(0,-1){57}}
\put(60,110){\vector(0,-1){30}}
\put(56, 109){\bf 2}
\put(50,60){\oval(20,20)[br]}
\put(50,40){\oval(20,20)[tl]}

\put(75,117){\line(0,-1){57}}
\put(75,110){\vector(0,-1){30}}
\put(70, 105){\bf 3}
\put(85,60){\oval(20,20)[bl]}
\put(85,40){\oval(20,20)[tr]}

\put(48,20){\line(1,0){17}}
\put(48,20){\vector(1,0){10}}
\put(50,15){$t \sim t_{\rm pin}$}

\end{picture}
\end{center}
\caption{Flowchart illustrating the three evolutionary scenarios (1, 2, 3).   The various boxes (A, B, C, D) describe different stages of evolution, with all scenarios beginning  in box (A), which represents the onset of the flare itself.  The boxes give the range of magnetic field perturbations $\delta B(t)$ to which they apply, the predominant nature of the neutron fluid (normal/superfluid), information on the interaction between the vortices and fluxtubes (pinning/cutting), information on `mass loading' (i.e. whether the full stellar inertia participates in the Alfv\'en oscillation, or just some fraction), and a comment on the likely damping mechanisms that apply.  \label{fig:flowchart}}
\end{figure}
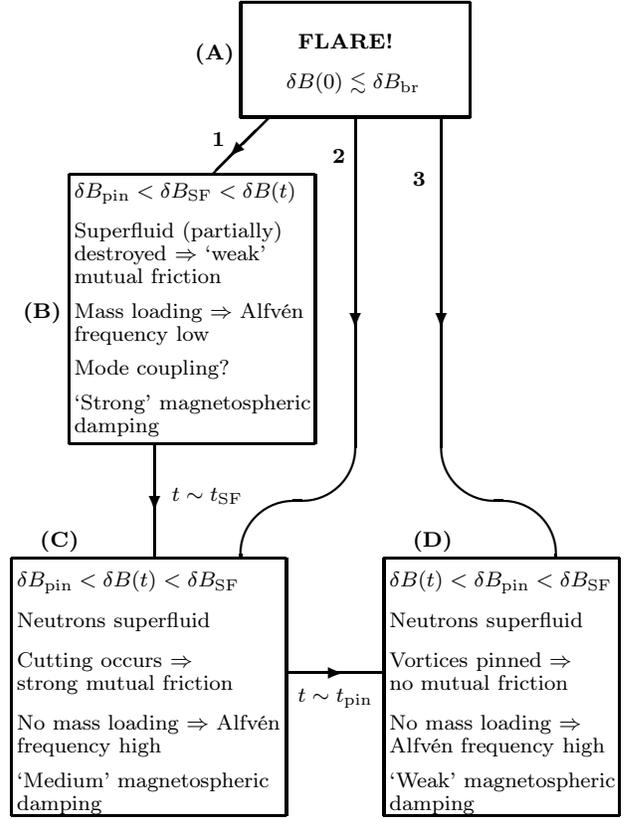

There are three key magnetic quantities that are of importance for magnetar flares.  These are (i) $\delta B(0)$, the initial value of the magnetic field perturbation, which we expect to be bounded by  $\delta B(0) \lesssim \delta B_{\rm br}$; (ii) $\delta B_{\rm SF}$, the critical perturbation above which superfluidity is destroyed; (iii) $\delta B_{\rm pin}$, the critical perturbation above which pinning between vortices and fluxtubes is broken, so that the dissipative process of fluxtube cutting occurs.  As given above, we expect $\delta B_{\rm pin} < \delta B_{\rm SF}$, so that there are \emph{three} possible orderings of the three quantities, depending upon the value of $\delta B(0)$ relative to the other two.  The magnetic field perturbation itself, $\delta B(t)$, will decay in time, generating  three distinct evolutionary pathways, illustrated in the flowchart of Figure \ref{fig:flowchart}, which we will describe below.

Before doing so, we will mention another important part of this story---the damping of the modes.  We consider damping in detail in Section \ref{sec:damping}.  There are two main sorts.  Modes can be damped by either \emph{external} or \emph{internal} mechanisms.  The external damping involves the launching of Alfv\'en waves into the magnetosphere, whose energy is eventually converted into an outward particle flux.  This acts on both Alfv\'en and crustal modes.  The efficiency of this process depends upon the fraction of the stellar surface which launches such waves.  We consider three possibilities: emission from  the entire stellar surface, emission only from a polar cap defined by the velocity-of-light cylinder, and emission only from a polar cap defined by an `Alfv\'en radius'.  The last of these seems to dominate, so is probably the most relevant.  The internal damping includes mutual friction, which can be due to the cutting of magnetic fluxtubes by neutron vortices as described above, or the more conventional scattering of electrons off neutron vortices.  This applies only to the Alfv\'en modes, as this mechanism requires relative motion between the interior neutron vortices and the charged component of the star.  A summary of our order-of-magnitude estimates of these various timescales is given in Table \ref{table:damping_times}.  Derivations of all of these results is given in Section (\ref{sec:damping}). 
\begin{table*}
 \begin{minipage}{165mm}
 \caption{Table of \emph{approximate} damping times, for both crustal and Alfv\'en modes.  All times in seconds, and relevant equation numbers are given.  The first column labels the type of mode.  The second, third and fourth columns give the damping times for magnetospheric damping, assuming emission over the full surface, over a polar cap defined by the velocity of light cylinder, and over a polar cap defined by the Alfv\'en radius, respectively.  The fifth and six columns give the damping times for mutual friction, assuming fluxtube cutting and electron scattering, respectively, relevant for Alfv\'en modes only.}
 \label{table:damping_times}
\begin{tabular}{@{}llllll}
 Damping mechanism: & \multicolumn{3}{c}{Magnetospheric damping} & \multicolumn{2}{c}{Mutual friction}\\ \hline
 & Full surface & Velocity of light & Alfv\'en radius & Fluxtube cutting & Electron \\
 & & cylinder& & & scattering\\
 \hline
 Crustal mode & $30$ [Eq.(\ref{tauC1})] & $10^6$ [Eq.(\ref{tauC2})]& $30 (\delta B/B)^{-2/3}$ [Eq.(\ref{tauC3})]& N/A & N/A \\
 Alfv\'en mode & $4$ [Eq.(\ref{taufull})]& $10^{10}$ [Eq.(\ref{tauL})] & $4 (\delta B/B)^{-4/3}$ [Eq.(\ref{tauA1})] 
 & $3 \left(\delta B/\delta B_{\rm pin}\right)^{3/2}$ [Eq.(\ref{taumf2})] & $10^3$ [Eq.(\ref{taumf_e})] \\ 
 \end{tabular}
\end{minipage}
\end{table*}

Having set the scene,  we can now describe the three distinct evolutionary pathways, illustrated in the flowchart of Figure \ref{fig:flowchart}, We begin by considering only Alfv\'en modes.  As will be explained below, the story for crustal modes is much simpler.  All pathways start in box A, which simply represents the cracking event itself, at time $t=0$.  

In pathway 1, we have, at early times, $ \delta B_{\rm pin} < \delta B_{\rm SF}  < \delta B(t)$.  In this case the initial perturbation of the star is large enough to destroy the superfluidity, as represented by box B.  This means that the neutrons and protons 
are tightly coupled, so that the Alfv\'en modes are `mass loaded', i.e. the full stellar inertia plays a role in determining the mode frequencies.  The Alfv\'en speed is then given by 
\be
v_\rA^2 = \frac{H_\rc B}{4\pi\rho} ,
\label{vA3}
\ee
with the corresponding mode frequencies being $\omega \sim v_{\rm A}/L$, where $L$ is the mode's wavelength (typically this would be comparable to the
stellar radius $R$). The tight coupling of neutrons to the charged component eliminates mutual friction as a damping mechanism, so in this case external damping
is likely to dominate, probably with a timescale given by the `Alfv\'en radius' column of Table \ref{table:damping_times} (i.e. equation (\ref{tauA1})).  We should also note that the large mode amplitudes
of this regime may mean that mode-mode coupling is important also, but we have no estimate of this.

After a time $\sim t_{\rm SF}$, of the order of the magnetospheric decay time of equation (\ref{tauA1}), the perturbation has decayed to the point such that $\delta B(t)$ is no longer large enough to destroy the superfluidity at any point in the star.  We then have $\delta B_{\rm pin} < \delta B(t) < \delta B_{\rm SF}$, as represented by box C.  The superfluid nature of the neutrons suppresses any microscopic coupling between them and the protons, such that the Alfv\'en modes are no longer mass loaded, i.e. only the inertia of the charged component plays a role in determining the mode frequency, and the Alfv\'en speed is given by equation (\ref{vA2}), which amounts to making the replacement $\rho \rightarrow \rho_\rp$ in equation (\ref{vA3}).   Simultaneously, cutting occurs between the magnetic fluxtubes and the (newly formed) neutron vortices, opening up mutual friction, a new and powerful mechanism for damping, probably dominated by the cutting itself, rather than the scattering of electrons off the fluxtubes, as indicated by the last two columns of the Table.  We can therefore expect the frequencies of Alfv\'en modes to \emph{increase} at $t \sim t_{\rm SF}$, by a factor of $\sim (\rho/\rho_\rp)^{1/2} \approx  x_{\rm p}^{-1/2} \approx 4$, while their damping time \emph{decreases} to a value given by equation (\ref{taumf2}), which is itself a function of $\delta B(t)$.

After a time $t_{\rm pin}$, of the order of the fluxtube cutting timescale of equation (\ref{taumf2}),  the perturbation will decay to the point where  $ \delta B(t) < \delta B_{\rm pin} <\delta B_{\rm SF}  $, as represented by box D.  Now the pinning has been re-established.  This will not have any effect on mode frequencies, but will shut-off the powerful damping mechanism of fluxtube cutting, so that the decay time of the Alfv\'en modes will increase, reverting again to a value determined by the launching of Alfv\'en waves into the magnetosphere, as per equation (\ref{tauA1}).

It is important to note that the transitions in damping and mass loading between the different regimes will not be perfectly sharp\footnote{We are grateful to the anonymous referee for  alerting us to this issue.}.  This is because the mode eigenfunctions for the axisymmetric ($m=0$) modes thought to be relevant to magnetar flares vanish at the origin, so there will always be a small volume around the origin where the velocity is less than any given critical value.  Also, such modes have a time-dependence proportional to $\sin\omega t$, so that there will be oscillation phases where the velocity perturbation of \emph{all} points in the star vanishes, so the velocity is again less than any given critical value.  This means that a star in our regime B will contains points at which the strong mutual friction/non-mass loaded conditions of regime C apply, while there will be points for a star in our regime C where the weak mutual friction conditions of regime D apply.  For stars where the perturbation $\delta B(t)$ is much greater than the relevant threshold ($\delta B_{\rm SF}$ or $\delta B_{\rm pin}$) these volumes/times will be short compared to the total stellar volume/oscillation period, but will have the effect of smoothing the transitions between the different regimes when the perturbation $\delta B(t)$ exceeds the critical value by only a small amount.  

Intuitively, given the strong damping that applies in Box (C), we might expect there to be some significant mutual friction damping even in the regime of box (B), due to superfluid at locations/times where the perturbation is small, giving a relatively smooth transition from (B) to (C).  Similarly, we might expect there to be a rather abrupt decrease in the damping rate in the transition (C) to (D), with the weak damping of (D) applying only when there are no times/locations where the velocity exceeds the critical value.

To confirm this intuition concerning damping rates, we have carried out some investigations of these effects using simple oscillator toy models.   We considered the transverse vibrations of a one-dimensional string with velocity $v(x,t)$, parameterised by a maximum amplitude $v_0$,  acted upon by a drag force proportional to $-\lambda_> v$ for $|v(x,t)|  > v_{\rm c}$, and proportional to $-\lambda_< v$ for $|v(x,t)| < v_{\rm c}$.  In the case where $\lambda_> / \lambda_< \gg 1$ , as describes the transition $C \rightarrow D$ above, we do indeed find that the transition is sharp, in the sense that even when the mode amplitude has fallen to just twice the critical value, (i.e.\   $v_0 = 2v_{\rm c}$), the damping rate is $85\%$ of the rate calculated assuming $\lambda = \lambda_>$ always.  In the case where $\lambda_> / \lambda_< \ll 1$, as describes the transition $B \rightarrow C$ above, we find a more gradual transition.  Setting $\lambda_> / \lambda_<  = 0.1$, we find that when the velocity is within a factor of $2$ of the critical value (again $v_0 = 2v_{\rm c}$), the damping rate is $17\%$ of the value it would have if  $\lambda = \lambda_>$ at all times.  

Given the strong nature of the damping due to fluxtube cutting, this level of damping should not be neglected in Box (B), which is why we use the 
term `weak mutual friction' there. Extrapolating the results of the toy model we can quantify the strength of mutual friction in Box (B). The
fluxtube cutting timescale entry in Table~\ref{table:damping_times} (which represents the mutual friction timescale in regime (C))
should be multiplied by a factor $10-100$ for $\delta B \approx (1.5-5) \delta B_{\rm SF} $ respectively.

We have also carried out some numerical experiments to gain more insight into the transition in mass loading, using a simpler toy model of a point oscillator (rather than a string), with mass $m = m_>$ for $|v(t)| > v_{\rm c}$, and mass $m = m_<$ for $|v(t)| < v_c$.    This represents the transition from the (mainly) mass-loaded state of regime (B) to the non-mass loaded regime  (C).  We chose the mass ratio  $m_< /m_> = x_{\rm p}$, equal to the proton mass fraction in the core, as is appropriate for such a transition.   The mode period was calculated as a function of the ratio of the mode amplitude $v_0$ to the critical velocity, ($v_0/v_{\rm c}$).  As was obviously to be expected, the mode frequency was found to increase by a factor $x_{\rm p}^{-1/2}$ as the system transitioned from the $v_0/v_{\rm c} \gg 1$ regime to the $v_0/v_{\rm c} < 1$ regime.    

Less obviously, we found a rather sharp transition from the higher frequency oscillation to the lower frequency one, occurring when $v_0$ was only slightly greater than $v_c$.  For example, for $v_0 \approx 1.4 v_c$,  the time spent in the $m_{<}$ state is a meagre $\approx 0.08$ fraction of
the total oscillation period.  We take this as evidence that the corresponding transition in a real star might be sharp and that the system in our Box (B)
is essentially mass-loaded.

We can now move on to the second of our evolutionary parthways.  In pathway 2 the initial value of the perturbation is assumed to be smaller than that for pathway 1, such that at early times $ \delta B_{\rm pin} < \delta B(t) < \delta B_{\rm SF}$, so that box C applies.  In this case,  the neutrons are superfluid from the outset, but at early times the perturbation is large enough to break the pinning, so that the powerful damping mechanism of fluxtube cutting is active.    This will shut off after a time $\sim t_{\rm pin}$ (of the order of the timescale of equation (\ref{taumf2})), when pinning is restored and the Alfv\'en mode decay time will lengthen to a value probably determined by the launching of Alfv\'en waves into the magnetosphere (equation (\ref{tauA1})), as represented by box D.

In pathway 3, the initial value of the perturbation is assumed to be small, such that $\delta B (t) < \delta B_{\rm pin} < \delta B_{\rm SF}  $ initially, and therefore for all later times also, so that box D applies at all times.  In this case, the neutrons are superfluid from the outset, and the vortices pinned to the fluxtubes, so there is no evolution in Alfv\'en mode frequencies or damping times; the modes are never mass-loaded, and the damping is always dominated by the launching of waves into the magnetosphere, as per equation (\ref{tauA1}).

The story for crustal modes is much simpler.  If $\delta B(t)$ is greater than $\delta B_{\rm SF}$, the crustal superfluidity will be destroyed.  The strong coupling between the neutron and protons (i.e. the crustal lattice) then ensures that the crustal modes are fully mass-loaded, so that $\omega \sim v_{\rm s}/L$, with the full (crustal) mass density appearing in the expression for $v_{\rm s}$ (as in equation~(\ref{vs}) above).  However, the situation is very similar in the case $\delta B(t) < \delta B_{\rm SF}$.  In this regime, the neutrons are superfluid, but the strong entrainment believed to apply to the crustal superfluid effectively couples the neutrons to the protons, so the dynamics remains essentially unchanged.  Meanwhile, the notion of a critical $\delta B_{\rm pin}$ does not apply for the crust, as the magnetic fluxtubes open out into `classical' (i.e. non-superconducting) flux bundles in the crust, so there is no notion of fluxtube-vortex cutting.  So, the crustal modes are, to a good approximation, unaffected by the transitions between the states B, C and D of the flow diagram that play such an important role for the Alfv\'en modes. For purely crustal modes, mutual friction is ineffective (because the crust segment of each vortex is not magnetised), 
so that the damping timescale is likely to be dominated by the launching of Alfv\'en waves into the magnetosphere, as per equation (\ref{tauC3}).

In summary, for  Alfv\'en modes, we have a rather rich set of possibilities.  Pathway 1 is represented by $A \rightarrow B \rightarrow C \rightarrow D$ in the flowchart, with changes occurring in both mode frequencies and damping times as the evolution proceeds.  Pathway 2 has the flow $A \rightarrow C \rightarrow D$, with no frequency changes, but a change in damping timescale.  Pathway 3 is the simplest, represented by $A \rightarrow  D$, with no sudden changes in any mode properties.  For crustal modes, there are no corresponding transitions in mode properties.

In Sections \ref{sec:dynamics} and \ref{sec:damping} we provide detailed calculations to estimate the various damping timescales that feature in the argument  sketched above, 
leaving our concluding discussion of these ideas to Section~\ref{sec:conclusions}.


\section{The dynamics of superfluid magnetars}
\label{sec:dynamics}

\subsection{Multifluid formalism}
\label{sec:multifluids}

The magnetohydrodynamics of magnetars can be formulated on the basis of a multifluid model for the stellar interior.
Mature magnetars older than 500 years or so are expected to be cold enough~\citep{HGA12} as to primarily consist of superfluid neutrons, superconducting
protons and electrons (the possibility of having exotic types of matter in the inner core is not considered here). An additional component is the solid lattice
in the crust which can be modelled as a positively charged elastic fluid. The equations of motion for this system are well known, see for example~\citet{SFmagnetar}.

There are two Euler equations, one for the neutrons and one for the ``protons'', the latter component is identified either with the actual protons
in the core or  with the crustal lattice (plus the electrons in both regions). 

In this paper we are exclusively interested in fluid motion (to be more precise, oscillations) which is accompanied by
negligible perturbations in the fluid density and pressure with respect to their background values. This category of oscillations includes 
the modes that are typically invoked as the most likely interpretation behind the QPOs seen in magnetar flares (i.e. axisymmetric axial parity modes,
and non-axisymmetric polar-parity modes, both of which are nearly incompressible).

For this kind of motion the linearised Euler equations in the stellar rotating frame take the form:
\be
\partial_t ( \bv_\rn + \en \bw ) + 2 \mathbf{\Omega} \times \bv_\rn  = \frac{1}{\rho_\rn} 
\left ( \bF_{\rm mf} + \bF_\rn \right )
\ee
\be
\partial_t ( \bv_\rp -\ep \bw ) + 2 \mathbf{\Omega} \times  \bv_\rp 
= \frac{1}{\rho_\rp} \left ( \bF_{\rm mag} + \bF_{\rm el} -\bF_{\rm mf} +\bF_{\rm visc} \right ) ,
\ee
where $\bv_\rx$ ($\rx = \rn, \rp$) is the fluid velocity with respect to the background rigid body rotation with angular
frequency $\mathbf{\Omega} = \Omega \mathbf{\hat{z}}$ (we ignore any small spin-lag that could be present between the various components),
$\bw = \bv_\rp -\bv_\rn$ is the proton-neutron velocity difference and $\rho_\rx$ is the individual fluid density.
The coupling due to the entrainment effect is encoded in the parameters $\ex$. These are related to the effective particle masses
$m^*_\rx$ as $\ex = 1 - m^*_\rx/m_\rx$~\citep{prix04}.

The proton fluid in the core experiences a magnetic force $\bF_{\rm mag}$ which represents the macroscopically averaged tension of the
quantised fluxtubes. A magnetic force $\bF_\rn$ is also exerted on the neutron superfluid in the core, as a result of the contribution of the entrained
neutrons to the fluxtube tension (see~\citet{supercon} for details). In the region of the crust $\bF_{\rm mag}$ is replaced by the usual Lorentz force and $\bF_{\rm el}$ is 
the elastic force exerted on the crustal lattice. The neutron-proton fluids are coupled through the mutual friction force $\bF_{\rm mf}$; this represents
any coupling between the fluids mediated through the neutron vortices. Finally, the proton fluid experiences a frictional force $\bF_{\rm visc}$  arising from shear and bulk viscosity. 

The additional dynamical equations are the continuity equations which for incompressible flow (as assumed here) take the form
$\nabla \cdot \bv_\rx = 0 $, and the magnetic induction equation 
\be
\partial_t \delta \bB = \nabla \times ( \bv_\rp \times \bB ) ,
\label{induction}
\ee
where $\delta \bB$ is the perturbed magnetic field.

We can re-write the pair of the Euler equations in an equivalent form in terms of the velocity difference $\bw$
and the average velocity $\bv = \xp \bv_\rp + (1-\xp) \bv_\rn$, where $\rho = \rho_\rn + \rho_\rp$ is the total density and 
$x_\rp = \rho_\rp / \rho$ is the ``proton'' fraction. 

The resulting `average' and `difference' Euler equations are, respectively,
\be
\partial_t \bv +  2\mathbf{\Omega} \times \bv = \frac{1}{\rho} \left ( \bF_{\rm mag} + \bF_\rn + \bF_{\rm el} \right ) 
\label{sumeuler}
\ee
and
\begin{multline} 
(1-\eb) \partial_t \bw +  2\mathbf{\Omega} \times \bw =
\\
= \frac{1}{\rho_\rp} \left ( \bF_{\rm mag}  +\bF_{\rm el} + \bF_{\rm visc} \right ) -\frac{1}{\rho_\rn} \bF_\rn - \frac{1}{x_\rp \rho_\rn} \bF_{\rm mf} ,
\label{diffeuler}
\end{multline}
where we have defined $\eb = \en+\ep $ and used the property $\rho_\rn \en = \rho_\rp \ep$~\citep{prix04}.

In this work we consider two idealised types of oscillations: (i) `Alfv\'en' modes, where the restoring force is predominantly magnetic, and which involve fluid motion in the bulk of the core; (ii) `crustal', where the restoring force is predominantly elastic, and which  are confined
 mainly in the crust.  Real magnetar oscillations may be some sort of hybrid, with properties
intermediate between these two extremes.

Our analysis will require mode solutions at a level of order of magnitude precision. For the modes we consider, the continuity equations are
trivially satisfied and we can safely ignore the Coriolis force term because mature magnetars are slow rotators. The magnetic force can be approximated as 
(see \citet{mendell91,supercon} for the exact expression for this force)
\be
F_{\rm mag} \sim \frac{H_\rc\, \delta B}{4\pi L} ,
\label{Fmag}
\ee 
where $L$ is the mode's characteristic lengthscale and $H_\rc \approx 10^{15}\,\mbox{G}$ is the (lower) critical field for superconductivity~\citep{tilley}. 
If $F_{\rm mag}$ is taken to be the magnetic force in the crust (or, more generally, the magnetic force in the absence of superconductivity) 
we can still use (\ref{Fmag}) after making the replacement $H_\rc \to B$. From the general expression for $\bF_\rn$~\citep{supercon} we can also deduce that
this force is smaller than $F_{\rm mag}$, 
\be
\frac{F_\rn}{F_{\rm mag}} \sim  x_\rp \frac{\rho}{m^*_\rp} \frac{\partial m^*_\rp}{\partial \rho}  \ll 1 ,
\ee 
where $m^*_\rp $ is the effective mass of the entrained protons. This is a slowly varying function of density, see for example~\citet{chamel08}.
Moreover, a plane-wave analysis as in~\citet{SFmagnetar} reveals that $\bF_\rn$ does not lead to any modification to the propagation speed
of Alfv\'en waves (see eqn~(\ref{vA2})).

For the elastic force we have
\be
\bF_{\rm el} = \mu \nabla^2 \mathbf{\xi}_\rp \quad \to \quad F_{\rm el} \sim \rho v_s^2 \frac{\xi_\rp}{L^2} ,
\ee
where $\mu$ is the crustal shear modulus, $v_s^2 = \mu/\rho$ is the shear speed, and 
$\mathbf{\xi}_\rp$ is the displacement of the crustal lattice (it is related to the strain in the crust as $\xi_\rp \sim \psi L$).
The displacement vector can be eliminated with the help of the induction equation~(\ref{induction}),
i.e. $\xi_\rp \sim L \delta B/B$, leading to
\be
 F_{\rm el} \sim \frac{\rho v_s^2}{L} \frac{\delta B}{B} .
\ee

The standard method for calculating modes of dissipative systems is to find a mode solution without initially 
including the frictional forces (in our case $\bF_{\rm mf}$ and $\bF_{\rm visc}$). The dissipative action of these forces is subsequently
accounted for by using the inviscid mode solution in suitable energy integral expressions. This will be our approach too in the 
calculation of the damping of magnetar oscillations (see Section~\ref{sec:damping}).

The mode property most relevant for our analysis is the relative proton-neutron velocity $\bw$. A simple inspection of
the difference Euler equation~(\ref{diffeuler}) reveals that this counter-moving degree of freedom is a generic property of
the modes considered here. From the same equation we estimate
\be
w \sim (1-\eb )^{-1}  \frac{F_{\rm restore}}{\omega x_\rp \rho} ,
\ee
where $\omega$ is the mode's angular frequency and $F_{\rm restore} = \{F_{\rm mag}, F_{\rm el} \}$ is the main restoring force.

Applying this formula to a global Alfv\'en mode (for which the entrainment prefactor is of order unity and 
$F_{\rm restore} = F_{\rm mag}$) we obtain  
\be
w \sim \frac{v_{\rm A}^2}{L\omega} \frac{\delta B}{B} , 
\ee
where we have used the Alfv\'en speed:
\be
v^2_\rA = \frac{H_\rc B}{4\pi \rho \xp} .
\ee
To the accuracy we are working, we can also approximate $v_{\rm A} \sim \omega L$ to obtain
\be
w \sim v_{\rm A} \frac{\delta B}{B} .
\label{wA1}
\ee
Inserting the explicit form of $v_{\rm A}$ and parameterising we finally obtain:
\be
w \sim 10^8 \left(\frac{H_{15} B_{15}}{x_5 \rho_{14}}\right)^{1/2} \frac{\delta B}{B}\, \mbox{cm} / \mbox{s} ,
\label{wA2}
\ee
where we have introduced the scalings $\rho_{14} = \rho/10^{14}\mbox{gr}\,\mbox{cm}^{-3}$,
$H_{15} = H_\rc/10^{15}\,\mbox{G}$, $x_5 = x_\rp/0.05$.

For crustal modes $F_{\rm restore} = F_{\rm el}$ and we have 
\be
w \sim (1-\eb)^{-1} \frac{v^2_s}{L x_\rp \omega} \frac{\delta B}{B} .
\ee 
In this case we have retained the entrainment factor; detailed calculations suggest that the effective mass of the entrained
superfluid neutrons could be much higher than the bare neutron mass, i.e. $m_\rn^* \gg m_\rn$~\citep{chamel05,chamel12}. 
We can then approximate (recall that $x_\rp \ll 1$ in the inner crust)
\be
1-\eb = 1- \frac{\en}{x_\rp} \approx  \frac{1}{x_\rp} \frac{m^*_\rn}{m_\rn} .
\ee
Then
\be
w \sim  \frac{m_\rn}{m^*_\rn} \frac{v^2_s}{L \omega} \frac{\delta B}{B} .
\label{wcrust1}
\ee
If we make the approximation $v_{\rm s} \sim \omega L$ we finally obtain
\be
w \sim  \frac{m_\rn}{m^*_\rn} v_{\rm s}  \frac{\delta B}{B} .
\label{wcrust2}
\ee
Inserting numerical values
\be
w \sim 7 \times 10^6 v_{{\rm s}, 8} \left(\frac{15}{m_\rn^* / m_\rn}\right)  \frac{\delta B}{B} \, ~ \mbox{cm}/\mbox{s} ,
\label{wcrust3}
\ee
where we have introduced the scaling $v_{{\rm s},8} = v_{\rm s}/$($10^8$ cm/s) (recall that $v_s$ stays almost uniform throughout the
crust).   The value of $m^*_\rn / m_\rn$ in neutron star crusts is discussed in~\citet{SFmagnetar}; according to their Figure 1, which is in turn based on the calculations of
\citet{chamel05,chamel12}, a value of $m^*_\rn / m_\rn \approx 15$ is typical throughout much of the crust.

The above expressions for the velocity lag $w$ will provide the key input for the calculations underpinning the evolutionary paths of Section~(\ref{sec:paths}).

Apart from these, we will also need an estimate for the mode's kinetic energy $E_{\rm mode}$ in order to compute damping timescales
in Section~\ref{sec:damping}. For our two-fluid system this energy receives contributions from both the comoving and countermoving
degrees of freedom (see for instance~\citet{2fmodes}),
\be
E_{\rm mode} = \frac{1}{2} \int dV \rho \left [ v^2 + x_\rp (1-x_\rp)  \left ( 1- \frac{\en}{x_\rp} \right ) w^2 \right ] .
\label{Emode1}
\ee
From the Euler equations it also follows (after eliminating $\bF_{\rm restore}$) that
\be
\bv \approx x_\rp (1-\eb) \bw .
\ee
An Alfv\'en-type mode would then represent a mainly counter-moving oscillation,
\be
\bv \approx x_\rp \frac{m^*_\rp}{m_\rp} \bw \ll \bw .
\ee
Note that a typical range for the entrained proton mass is $0.3 <  m^*_\rp /m_\rp < 0.8$~\citep{chamel08}.
The mode energy (\ref{Emode1}) is dominated by the second, countermoving, term. We can easily obtain
\be
E_{\rm mode} \approx \frac{1}{2} \int dV \rho  x_\rp   \frac{m^*_\rp}{m_\rp}  w^2 .
\label{Emode2}
\ee
In the case of crustal modes, and as a result of the strong entrainment coupling, we find a dominantly comoving character:
\be
\bv \approx \frac{m^*_\rn}{m_\rn} \bw
\ee
This time the mode energy is dominated by the first term, and the energy formula becomes essentially that of
a single-component system,
\be
E_{\rm mode} \approx \frac{1}{2} \int dV \rho  v^2 ,
\label{Emode3}
\ee
where, for a crustal mode, the integral should be taken over the region of the crust.


\subsection{Destruction of superfluidity}
\label{sec:destruction}

In a recent publication,~\citet{GK13}, have argued that a sufficiently large velocity difference between the superfluid and non-superfluid components 
can destroy the superfluidity in neutron stars -- a phenomenon also seen in laboratory systems. 
More accurately, the superfluid energy gap $\Delta$ is not only a function of temperature $T$, but is also a function of the relative velocity
between the superfluid and the normal components. The gap is suppressed when this relative motion increases. 
In our case the normal component is identified with the electrons which in turn are nearly comoving with the proton fluid. Therefore the relevant relative velocity is
$\bw$ and $\Delta = \Delta (T,w)$.

The main result of~\citet{GK13} is their equation (12) which gives an expression for the critical relative velocity $w_{\rm SF}$  between the
superfluid and normal components 
above which superfluidity is destroyed:
\be
w_{\rm SF} = 10^7  \left(\frac{\Delta(T,0)}{10^9 \rm K}\right) \left(\frac{\rho_0}{\rho}\right)^{1/3} \, ~ \mbox{cm}/\mbox{s} ,
\label{wSF}
\ee
where $\rho_0 = 0.16$ fm$^{-3}$ is the nuclear density and the neutron pairing gap has been normalized to a canonical
value. 

Clearly, it is interesting to compare this velocity with the $w$ defined above, which gives the superfluid--charged particle velocity difference for an Alfv\'en-type oscillation.
Comparison with equation (\ref{wA2}) above shows that, for Alfv\'en modes, we can expect the actual velocity difference $w$ to exceed the critical value for destroying the superfluidity for $\delta B / B \sim 0.1$, indicating that the larger flare events may well push the star over this critical threshold.  Comparing instead with equation  (\ref{wcrust3}) above we see that, for crustal modes, somewhat larger perturbations in the magnetic field are required to break the superfluidity, with $\delta B / B \sim 1$.  
This is large, but, given the order-of-magnitude nature of our estimates, the possibility of crustal modes also breaking the superfluidity cannot be excluded.

As discussed in Section \ref{sec:key_thresholds} above, for the purposes of describing the evolution of a magnetar flare, it is convenient to recast our results in terms of $\delta B / B$, the critical (fractional) perturbation in magnetic field above which superfluidity is destroyed.  To do so, for Alfv\'en modes, we set $w = w_{\rm SF}$ in equation (\ref{wA1}) above and invert to give
\be
\frac{\delta B_{\rm SF}}{B} \sim \frac{w_{\rm SF}}{v_{\rm A}} .
\ee
Parameterising:
\be
\frac{\delta B_{\rm SF}}{B} \sim 0.08 \, \left(\frac{x_5\rho_{14}}{H_{15}B_{15}}\right)^{1/2} w_{{\rm SF}, 7} ,
\ee  
where $w_{{\rm SF}, 7}= w_{\rm SF}/10^7 \,\mbox{cm}\,\mbox{s}^{-1}$.
For crustal modes, we similarly set $w=w_{\rm SF}$ in equation (\ref{wcrust2}) to give
\be
\frac{\delta B_{\rm SF}}{B} \sim \frac{m_\rn^*}{m_\rn}    \frac{w_{\rm SF}}{v_{\rm A}} .
\ee
Parameterising:
\be
\frac{\delta B_{\rm SF}}{B} \sim     \left(\frac{m_\rn^*/m_\rn}{15}\right) \frac{w_{{\rm SF}, 7}}{v_{{\rm s}, 8}} .
\ee

\section{Damping of magnetar oscillations}
\label{sec:damping}

\subsection{Magnetospheric damping: formalism}

A stellar oscillation that involves motion of the stellar surface will also ``shake'' the magnetic field lines, launching
Alfv\'en waves in the magnetosphere. These waves effectively remove energy from the oscillation and dissipate it as they
propagate out across the magnetosphere. This process could be relevant for dissipating magnetar oscillations excited during 
flare events and it is the purpose of this section to assess its efficiency.

Following the analysis of \citet{HL00} we consider a typical fluid displacement $\mathbf{\xi}_0$ on the stellar surface;
the induced Alfv\'en wave will travel a distance $\sim c/\omega$ in the magnetosphere in one oscillation period. 
The resulting wave amplitude will be $\delta B \sim B \omega \xi_0/c $ -- this is much smaller than
magnetic field perturbation $\sim \xi_0 B/L$ induced in the stellar interior (see eqn.~(\ref{induction})).

The radiated power is given by the following Poynting flux surface integral
\be
P_\rA \approx 
\frac{c}{8\pi} \int dS \frac{\omega^2}{c^2} | \mathbf{\hat{B}} \cdot \mathbf{\hat{r}} | 
\left | \mathbf{\xi}_0 \times \bB \right |^2 ,
\label{PA0}
\ee
where $\mathbf{\hat{B}}, \mathbf{\hat{r}}$ are unit vectors along the magnetic field and the radial direction, respectively.
This can be approximated as
\be
P_\rA \sim  4\pi R^2 c \frac{(\delta B)^2}{8\pi} \sim \frac{1}{2c} (\omega \xi_0 B R)^2 .
\label{PA1}
\ee

These formulae allow for emission over the entire stellar surface.  However, this is not necessarily a good assumption.   
Firstly, it only really makes sense to talk of Alfv\'en waves on field lines whose total length is significantly longer than the wavelength of the Alfv\'en waves.  Secondly, as discussed in \citet{TB98}, it is likely that this flux only represents an energy loss for those field lines which are \emph{open}.  
The idea is that Alfv\'en  waves on open field lines propagate out far from the star and at some point become non-linear and radiate their energy away as electromagnetic waves and/or particles. 
The upshot of this effect is to suppress the flux as calculated by an all-surface integral by a factor, so as to allow for energy loss only from the \emph{polar cap}, i.e. from on a patch of angular radius $\theta_{\rm cap}$, centred on the magnetic axis. defined to be the surface region from which open field lines emanate.  In the following calculations, we will make the small-angle approximation $\sin\theta_{\rm cap} \approx \theta_{\rm cap}$, although there are situations where the polar cap can occupy a significant fraction of the stellar surface.  Given the order-of-magnitude accuracy of other parts of our calculation, such an approximation is acceptable.

It is useful to relate $\theta_{\rm cap}$ to the (cylindrical) radius of the last closed field line.  To do so, note that a dipolar magnetic field line is given by the polar 
relation $r(\theta)$:
\be
\frac{\sin^2\theta}{r(\theta)} = \frac{1}{r(\pi/2)} ,
\ee
assuming a dipole axis along $\theta = 0$.  Suppose the last closed field line cuts the equatorial $\theta= \pi/2$ plane at a radius $R_{\rm c}$.   Then this field line is defined by the equation
\be
\frac{\sin^2\theta}{r(\theta)} = \frac{1}{R_{\rm c}} .
\ee
It follows that this field line cuts the stellar surface $r=R$ at  $\theta = \theta_{\rm cap}$ given by
$\sin\theta_{\rm cap} = \sqrt{R/R_{\rm c}}$.  We can therefore approximate
\be
\label{theta_cap}
\theta_{\rm cap} \approx \sqrt{\frac{R}{R_{\rm c}}} \lesssim 1 .
\ee
The precise value of $R_{\rm c}$ then depends upon which physical mechanism is responsible for determining the last closed field line.  This argument will hold up to geometric factors of order unity for non-aligned dipoles, i.e. dipole whose symmetry axes do not lie along $\theta = 0$.  Note that the polar cap will always be centred on the magnetic dipole axis, not the rotation axis.

It is standard practice in modelling neutron star magnetospheres (especially those of radio pulsars) to place the last open field line at the location (or thereabouts) where the rotational velocity of the
field lines equals the speed of light, i.e. the location of the light cylinder $R_\rL = c/\Omega$ (e.g. \citet{GJ69}).  We then have, using eqn.~(\ref{theta_cap}),
\be
\label{thetaL}
\theta_\rL \approx \sqrt{\frac{2\pi R}{cP}} \approx 0.015 \, R_6^{1/2} \left ( \frac{P}{1\,\mbox{s}} \right)^{-1/2} .
\ee

In addition to the finite rotation of the system there is another mechanism that can cause the lines to become open, active only in dynamically perturbed (rather than rigidly-rotating) magnetospheres. This mechanism has been discussed by~\citet{TB98}.  Consider a ``tube" of Alfv\'en waves propagating out along the field lines that originate from the polar cap.  The magnetic pressure from the (background) dipolar field scales as $B^2$, which is a steeply decaying function of distance from the star ($\sim r^{-6}$).  The Alfv\'en wave pressure is a less steep function (recall that   $\delta B \sim B^{1/2}$ for the travelling Alfv\'en wave~\citep{blaes89}).   \citet{TB98} argue that there then  exists a critical `Alfv\'en radius' at which the Alfv\'en stresses dominate the magnetic dipole ones, thus opening up the field lines. \citet{TB98} calculate a polar cap angular radius
\be
\theta_\rA \approx \left(\frac{\delta B}{B}\right)^{1/3} ,
\label{thetaA1}
\ee
where both $B$ and $\delta B$ are to be evaluated at the stellar surface.  The dependence on $\delta B$ makes this cap radius dependent upon the amplitude of the perturbation. 

To gain some insight,  we can easily calculate the ranges in  polar cap radius corresponding to the various stages in evolution of Alfv\'en modes discussed in Section \ref{sec:paths}.   For the evolutionary stage represented by box B in the flowchart of Figure \ref{fig:flowchart} we can combine equations (\ref{dBbr}), (\ref{dBsf_alfven})  and (\ref{thetaA1}) to obtain
\be
\delta B_{\rm SF} <  \delta B < \delta B_{\rm br} \iff 0.4 \lesssim \theta_\rA \lesssim 1 .
\label{thetaA2}
\ee
For the evolutionary stage represented by box C in the flowchart  we can also use equation (\ref{dBpin}) to obtain
\be
\delta B_{\rm pin} < \delta B < \delta B_{\rm SF} \iff 0.16 \lesssim \theta_\rA \lesssim  0.4 .
\label{thetaA3}
\ee
For crustal modes, the threshold $\delta B_{\rm pin}$ is irrelevant, while the value of $\delta B_{\rm SF}$ is given by equation (\ref{dBsf_crust}), which, when  combined with equation (\ref{thetaA1}) gives a corresponding polar cap radius of order unity.

Comparing the above results with equation (\ref{thetaL}) above, we see that for both Alfv\'en and crustal modes, the polar cap radius in the present case can be much larger than the one in the light cylinder model.  More quantitatively, this opening of field lines due to Alfv\'en pressure will be  the determining factor in fixing $\theta_{\rm cap}$  when $\theta_\rA > \theta_\rL$.   Combining equations (\ref{thetaL}) and (\ref{thetaA3}) we see that this is the case when $P > 0.02 R_6^{1/2}\,\mbox{s}$. 
This is clearly well satisfied for magnetars.

There is another modification of the Poynting flux  due to the open field lines that has to do with the nature of the oscillation mode itself. 
For most classes of modes we would expect that the fluid displacement in the cap would be $\xi_{\rm cap} \approx \xi_0$. 
However, for magnetic Alfv\'en modes, the mode eigenfunction will vanish on the magnetic axis, and we would expect $\xi_{\rm cap} \approx \xi_0 \theta$ in the vicinity of the magnetic axis.  We will therefore estimate the Alfv\'en flux as
\be
\label{PA2}
P_{\rm A} \approx \frac{R^2}{4c} \theta_{\rm cap}^2 (\omega \xi_{\rm cap} B)^2 ,
\ee
with
\bear
&& \xi_{\rm cap} \approx \xi_0 \theta_{\rm cap} \qquad \mbox{for Alfv\'en modes} ,
\label{xiA}
\\
&& \xi_{\rm cap} \approx \xi_0 \qquad \qquad \mbox{for other modes} .
\label{xicrust}
\eear

We will proceed by first evaluating the Poynting fluxes and magnetospheric damping times by performing a naive full-surface integral for crustal modes and Alfv\'en modes, making use of equation (\ref{PA1}). The damping timescale $\tau_\rA$ will be calculated with the help of the standard formula
\be
\tau_\rA = \frac{E_{\rm mode}}{P_\rA} ,
\ee
where $E_{\rm mode}$ is the mode energy. 
We will then calculate damping times in the case where the location of the magnetospheric light cylinder defines the last closed field line,  making use of equations (\ref{PA2}) and (\ref{thetaL}).  We will then repeat this procedure for the case where the Alfv\'en radius  defines the last closed field line, again making use of equation (\ref{PA2}), this time combined with equation (\ref{thetaA1}).
Of course, this is just a rough estimate; a more accurate calculation would solve for the integral in equation (\ref{PA0}) over the appropriate range in $\theta$, inserting the correct inclined-dipole geometry and mode eigenfunction. This, however, seems an overkill given that we are looking for 
approximate indications of what are the important damping mechanisms for magnetar oscillations.

In the calculations that follow, we first present the damping assuming emission over the full surface, which we are then able to easily modify to allow for emission only over a polar cap, defined either by the last closed field line or balance between the magnetic and Alfv\'en pressures.  It is these last two results that we consider to be the realistic ones, so it is the \emph{shorter} of these two that we expect to be relevant in a magnetar.  (This in fact proves to be the  result obtained by balancing the magnetic and Alfv\'en pressures; compare columns 3 and 4 of Table \ref{table:damping_times}.)


\subsection{Magnetospheric damping: Alfv\'en modes}
\label{sec:alfven}

As we have already discussed, the Alfv\'en-type modes are assumed to be global oscillations, extending over the bulk of the 
neutron star core. The mode energy has been given in (\ref{Emode2}). For the purposes of the present calculation we can approximate
that expression by assuming that $\xi_0$ is comparable to the fluid displacements in the core, that is, $w \sim \omega \xi_0$. Then
\be
\label{eq:E_mode_alfven}
E_{\rm mode} \sim \frac{1}{2} x_\rp \rho \omega^2 \xi^2_0 V_* ,
\ee 
where $V_*$ is the stellar volume.

Combining the mode energy with the Poynting flux (\ref{PA1}) for the entire surface we obtain the following estimate for the damping 
timescale\footnote{We note that a more rigorous calculation for the damping of axial and axisymmetric Alfv\'en modes in a uniform density fluid sphere coupled with the Poynting
flux~(\ref{PA0})  leads to a similar result.}:
\be
\label{taufull} 
\tau_\rA \sim \frac{x_\rp Mc}{B^2 R^2} \, \sim \, 4 x_5 M_{1.4} R^{-2}_6 B^{-2}_{15}\,~\mbox{s} ,
\ee 
where $M_{1.4} = M/1.4\,M_\odot$.

We will now re-evaluate the Alfv\'en fluxes assuming energy losses only along open field lines, assuming that the last closed field line is defined by the velocity-of-light cylinder.  Using equations (\ref{PA1}) and (\ref{xiA}) we obtain new, longer, damping times.  The flux $P_{\rm A}$ now contains a factor of $\theta^4_\rL$:
\be
P_{\rm A} \sim \pi^2 \frac{R^4}{c^3 P^2} (\omega \xi_0 B)^2 .
\ee
The mode energy is the same as before, and we thus obtain the (much longer) timescale 
\be
\tau_{A} \sim  \frac{x_\rp}{2\pi^2} \frac{Mc^3 P^2}{R^4 B^2}
\approx 10^{10}  x_5 \left ( \frac{P}{10\,\mbox{s}} \right )^2  \frac{M_{1.4}}{R_6^4 B_{15}^2}\,~\mbox{s} .
\label{tauL}
\ee

For the case where the polar cap radius is determined by the magnetic pressure we can again use 
(\ref{PA1}) and (\ref{xiA}) with $\theta_{\rm cap} = \theta_\rA$: 
\be
P_{\rm A} \sim \left ( \frac{\delta B}{B} \right )^{4/3} \frac{R^2}{4c} \left ( \omega \xi_0 B \right )^2 .
\ee
This time the  damping timescale is 
\be
\tau_{\rm A} \sim  \left ( \frac{\delta B}{B} \right )^{-4/3}  \frac{x_\rp M c}{B^2 R^2}
\sim 4   \left ( \frac{\delta B}{B} \right )^{-4/3} \frac{x_5 M_{1.4}}{B_{15}^2R_6^2} \,~\mbox{s} .
\label{tauA1}
\ee
Using the polar cap sizes of eqns.~(\ref{thetaA2}) and (\ref{thetaA3}) we find
that in the range $\delta B_{\rm SF} < \delta B < \delta B_{\rm br}$,
\be
\tau_\rA \sim ( 0.05-  150 )\, x_5 M_{1.4} R^{-2}_6 B^{-2}_{15}\,~\mbox{s} ,
\label{tauA2}
\ee
and 
\be
\tau \sim  ( 150 -  5 \times 10^3 )\,  x_5 M_{1.4} R^{-2}_6 B^{-2}_{15}\,~\mbox{s} ,
\label{tauA3}
\ee
in the range $\delta B_{\rm pin} < \delta B < \delta B_{\rm SF}$.   (Note that these timescales, in common with some others that will follow below, depend upon the size of the perturbation itself.  This non-exponential damping  is characteristic of non-linear damping processes.  Such timescales are physically meaningful, having the interpretation of the time it takes for the mode to decay by a factor of order unity.)

As expected these timescales lie somewhere between the full surface and light cylinder results, eqns.~(\ref{taufull}) and (\ref{tauL})
respectively. It is also clear that the full surface timescale (\ref{taufull}) and the (perhaps more  realistic) polar cap-modified timescale
(\ref{tauA2}) are sufficiently short as to play a role in the evolutionary pathways discussed earlier in Section~\ref{sec:paths}. On the other hand 
the light-cylinder timescale (\ref{tauL}) is too long to have any bearing on our analysis.  



\subsection{Magnetospheric damping: crustal modes} \label{sect:md:cm}

The second class of oscillations under consideration is that of crustal elastic modes. For these modes the fluid motion is
confined inside the crust. 

As before, we will compute the damping timescale $\tau_\rA$ by first using the full-surface Poynting flux (\ref{PA1}), making no effort to account for the polar 
cap suppression factor. The mode energy is given by eqn.~(\ref{Emode3}) and we can approximate that by assuming that $v \sim \omega \xi_0$. Then
\be
E_{\rm mode} \approx \frac{1}{2} \rho (\omega \xi_0)^2 V_{\rm crust} ,
\ee
where $V_{\rm crust}$ is the volume of the crust. 

We obtain
\be
\tau_\rA \sim c \rho \frac{V_{\rm crust}}{B^2 R^2} \approx 3  c \frac{M \Delta}{B^2 R^3} ,
\ee
where $\Delta$ is the thickness of the crust (we have also used $\rho \approx 3M/4\pi R^3$ for the density in the inner crust). 
Choosing a typical thickness $\Delta = 0.1 R$ we find\footnote{This result is of the same order of magnitude as the more rigorous timescale
calculated for axial and axisymmetric crustal modes of a uniform crust coupled with the flux (\ref{PA0}).}
\be
\tau_\rA \sim 0.3  \frac{Mc}{B^2 R^2} ,
\ee
or equivalently 
\be
\label{tauC1}
\tau_\rA \approx 26 \, M_{1.4} R^{-2}_6 B_{15}^{-2}\,~\mbox{s} .
\ee

If the magnetospheric damping is limited to the open field lines then we have the revised timescale (through eqn.~(\ref{PA1}) and the light-cylinder correction factor 
$\theta_\rL^2$): 
\be
\tau_\rA \sim 0.05 \frac{M c^2 P}{ B^2 R^3} \approx 10^6\, \left ( \frac{P}{10\,\mbox{s}} \right )  M_{1.4} R^{-3}_6 B_{15}^{-2}\,~\mbox{s} .
\label{tauC2}
\ee

Lastly, if the open line region is determined by the magnetic pressure the timescale is modified by a factor $\theta_\rA^{_2}$:
\be
\tau_\rA \sim 0.3 \left ( \frac{\delta B}{B}\right )^{-2/3} \frac{Mc}{B^2 R^2}
\sim 30   \left ( \frac{\delta B}{B} \right )^{-2/3} \frac{M_{1.4}}{B_{15}^2R_6^2} \,~\mbox{s} .
\label{tauC3}
\ee
For  the amplitude range $\delta B_{\rm SF} < \delta B < \delta B_{\rm br}$ we have
\be
\tau_{\rm A} \sim ( 26 -  160) \, M_{1.4} R^{-2}_6 B_{15}^{-2}\,~\mbox{s} ,
\ee
while for the range $\delta B_{\rm pin} < \delta B < \delta B_{\rm SF}$,
\be
\tau_{\rm A} \sim ( 160 -  10^3 ) \, M_{1.4} R^{-2}_6 B_{15}^{-2}\,~\mbox{s} .
\ee 
Apart from the light-cylinder result, all other damping timescales are rather short and are potentially relevant in the analysis of  the evolution of crustal modes.


\subsection{Internal damping: friction due to fluxtube cutting}
\label{sec:cutting}

The co-existing and possibly intersecting neutron vortices and proton fluxtubes in the outer core of neutron stars
can lead to a substantial mutual friction force $\bF_{\rm mf}$ between the fluids. The basic reason behind the
strong vortex-fluxtube interaction is their intrinsic (mesoscopic) magnetic fields. The force at a single intersection site 
is given by the ratio of the local magnetic interaction energy $E_{\rm int}$ to the range of the magnetic forces, which can be taken to be
the London penetration length $\Lambda$. That is, after ignoring geometric factors of order unity \citep{ruderman98},
\be
F_{\rm int} \approx \frac{E_{\rm int}}{\Lambda} \approx \Lambda^2 B_\rn B_\rp ,
\ee
where $B_\rn, B_\rp$ are the magnetic fields carried by individual vortices and fluxtubes, respectively. The force exerted 
per unit vortex length is
\be
f_{\rm pin} \approx \frac{F_{\rm int}}{d_\rp} ,
\ee 
where  $d_\rp \approx 95\, B_{15}^{-1/2}\,\mbox{fm} $ is the typical distance between fluxtubes.

In the unperturbed system the vortex array is expected to be pinned to the much more numerous fluxtubes, but an oscillation with
a sufficiently large amplitude may cause large scale vortex unpinning. In that case the vortices will be able to 
move by ``cutting'' through the fluxtubes. In both cases $f_{\rm pin}$ represents the pinning/interaction force.

The process of cutting is dissipative: part of the fluid's kinetic energy is transformed into short-wavelength ``kelvon'' excitations 
induced along each vortex (see~\citet{link03} for a discussion). Each kelvon has an effective mass $\mu$, wavenumber $k$ and
energy $E_k = \hbar^2 k^2/2\mu$. If the relative fluxtube-vortex velocity is    
\be
\bu_{\rm rel} =  \bu_\rp - \bu_\rn ,
\ee
then the interaction at each vortex-fluxtube intersection lasts a time interval $t_{\rm int} \sim \Lambda_\rp/u_{\rm rel}$.
From the uncertainty principle we also get $E_k \approx \hbar/t_{\rm int} $ and this leads to
\be
k \approx \left ( \frac{2\mu}{\hbar \Lambda} u_{\rm rel} \right )^{1/2} 
\equiv \frac{1}{\Lambda} \left ( \frac{u_{\rm rel}}{v_\Lambda} \right )^{1/2} ,
\ee
where we have defined
\be
v_\Lambda = \hbar/2\mu\Lambda \approx 10^9\, \mbox{cm}/\mbox{s} .
\ee

This qualitative analysis is in agreement with the more detailed discussion of~\citet{link03}.  Proceeding, we can note that for unpinning to occur, $u_{\rm rel}$ should 
exceed the fluid velocity lag threshold $w_{\rm pin}$ above which the Magnus force exceeds the pinning force and drives vortex 
unpinning\footnote{In a state of pinning $ \bu_\rn \approx \bu_\rp \approx \bv_\rp$ and hence the velocity difference appearing in the Magnus force is $\bu_\rn - \bv_\rn \approx \bw $}. This critical lag is~\citep{link03}:
\be
w_{\rm pin} \sim \frac{f_{\rm pin}}{\rho_\rn \kappa \varpi} \quad \to \quad 
w_{\rm pin} \sim 5 \times 10^5\,\rho_{14}^{-1} B_{15}^{1/2} \, ~\mbox{cm}/\mbox{s} ,
\label{wpin}
\ee
where $\kappa = h/2m \approx 2 \times 10^{-3}\, \mbox{cm}^2/\mbox{s} $ is the quantum of circulation for
individual vortices (we have also set $\varpi = 10^6\,\mbox{cm}$ as a representative value for the cylindrical radial
coordinate in the outer core).  Comparing with equation (\ref{wA2}) giving our estimates of the value of fluids' relative velocity excited by the flare event, we see that larger flares can easily satisfy $w_{\rm pin} < u_{\rm rel}$, so it would seem that the vortices can indeed unpin.

However, Link's calculation treats the vortex--fluxtube  intersections as widely separated in relation with the kelvon wavelength, thus ensuring that the excitations do not add up coherently. Mathematically this requirement means that $k d_\rp \gg 1$, from which we obtain a lower limit on the relative velocity, above which we can trust the damping calculation: 
\be
u_{\rm low} \approx 6.5 \times 10^8\, B_{15}\,~ \mbox{cm}/\mbox{s} .
\ee  
Link's fluxtube cutting model would then be valid for $u_{\rm rel} \gg u_{\rm low}$.   We can see that for magnetar-strong magnetic fields $u_{\rm low}$ is quite large and we should expect 
$u_{\rm low} \gg w_{\rm pin}$. We should also expect that for any reasonable oscillation amplitude we would have
$u_{\rm rel} \ll u_{\rm low} $.   In other words, a magnetar oscillation capable of forcing vortex unpinning in the first place would be expected to lead to a vortex-fluxtube relative velocity somewhere in the range
\be
w_{\rm pin} \lesssim u_{\rm rel} \lesssim u_{\rm low} .
\ee
Unfortunately, this range lies \textit{outside} the validity of Link's fluxtube cutting model. For lack of a better alternative
we opt for ``abusing'' Link's analysis and \emph{assume} that it remains reliable in the above kinematical range, at least as 
an order of magnitude estimate. This assumption is clearly debatable and this issue should be addressed by more detailed future work.
We can note, however, that as long as the kelvon wavelength is microscopically short, $k R \gg 1$, (which is always the case in
the relevant parameter range) some dissipation should take place.

The energy penalty for cutting through a single vortex-fluxtube intersection is~\citep{link03}:
\be
\Delta E = \frac{2}{\pi} \frac{F^2_{\rm int}}{\rho_\rn \kappa} ( v_\Lambda u_{\rm rel} )^{-1/2} .
\ee
The resulting dissipation rate per unit volume can then be written as
\be
\cE_{\rm cut} = \frac{\cN_\rn u_{\rm rel}}{d_\rp^2} \Delta E ,
\ee
where $\cN_\rn \approx 2\Omega /\kappa$ is the number of vortices per unit area. Combining the previous expressions,
\be
\dot{\cE}_{\rm cut} = \frac{4 \Omega  f^2_{\rm pin}}{\pi \rho_\rn \kappa^2} \left ( \frac{u_{\rm rel}}{v_\Lambda} \right )^{1/2} .
\label{dEcut}
\ee

The next step is to relate the vortex/fluxtube velocities with the fluid ones. Based on the high conductivity of the system
we can safely assume that the fluxtubes move together with the proton fluid, i.e.  $\bu_\rp \approx \bv_\rp$. It is not so easy to make 
a similar statement for the vortices and the neutrons. The simplest assumption is to set $\bu_\rn \approx \bv_\rn$; this could be accurate
in the range $u_{\rm rel} \gg w_{\rm pin}$ where cutting is strong and the relative motion of vortices is almost that of free vortices.
With these identifications we then have: 
\be
\bu_{\rm rel} = \bw .
\ee

The dissipation rate (\ref{dEcut}) can be also expressed as the work done by the mutual friction force $\bF_{\rm mf}$.
To this end we first consider the drag force per unit vortex length of the general form
\be
\mathbf{f}_{\rm D} = \rho_\rn \kappa \cR \bu_{\rm rel} .
\label{fv}
\ee 
The (dimensionless) drag coefficient $\cR$ is velocity-dependent,
\be
\cR = \cR_0 \left ( \frac{u_{\rm rel}}{v_\Lambda} \right )^{-3/2} ,
\ee
and $\cR_0$ is a constant parameter (the $-3/2$ exponent is dictated by eqn.~(\ref{work}) below). 

We can subsequently define the mutual friction force (per unit volume):
\be
\bF_{\rm mf} = \cN_\rn \mathbf{f}_{\rm D} \quad \to \quad  \bF_{\rm mf} = 2\Omega  \rho_\rn \cR_0  
\left ( \frac{u_{\rm rel}}{v_\Lambda} \right )^{-3/2}  \bu_{\rm rel} .
\ee
We then have
\be
\dot{\cE}_{\rm cut} = \bF_{\rm mf} \cdot \bu_{\rm rel} ,
\label{work}
\ee
which fixes
\be
\cR_0 =  \frac{2}{\pi} \left( \frac{f_{\rm pin}}{\rho_\rn \kappa v_\Lambda} \right )^2 \quad \to
\quad \cR_0 \approx 10^{-7} B_{15} .
\ee

The above equations allow the calculation of a mutual friction damping rate
\be
\dot{E}_{\rm fm} = \int dV \bF_{\rm mf} \cdot \bu_{\rm rel} \approx \int dV \bF_{\rm mf} \cdot \bw .
\ee
This becomes (assuming $\rho \approx \mbox{constant}$)
\be
\dot{E}_{\rm fm} \approx \frac{16\Omega}{\rho \kappa^2} \frac{f^2_{\rm pin}}{v_\Lambda^{1/2}} \int_{R_{\rm in}}^R dr r^2 w^{1/2} ,
\ee
where the radial integral is taken over the region of the outer core.

We can apply the above analysis to the case of magnetar  Alfv\'en oscillations. As we have seen in Section~(\ref{sec:multifluids}) we can approximate
the mode's kinetic energy as (given the other uncertainties in this calculation we can ignore the entrainment correction):
\be
E_{\rm mode}  \approx \frac{1}{2} \rho_\rp \int dV w^2 = 2\pi x_\rp \rho \int_0^R dr r^2 w^2 .
\ee
The mutual friction damping timescale can be estimated with the help of the standard formula:
\be
\tau_{\rm mf} = \frac{E_{\rm mode}}{| \dot{E}_{\rm mf}|} .
\ee
From this
\be
\tau_{\rm mf} \approx \frac{\pi \kappa^2 \rho^2 x_\rp}{4\Omega} \frac{v_\Lambda^2}{f_{\rm pin}^2}
\frac{\int_0^R dr r^2 w^2}{\int^R_{R_{\rm in}} dr r^2 w^{1/2}} .
\ee
Given the order-of-magnitude precision of our calculation we can assume $w \approx \mbox{uniform}$
and arrive at the simpler result
\be
\tau_{\rm mf} \sim \frac{\pi x_\rp}{4\Omega} \frac{v_\Lambda^{1/2}}{w^2_{\rm pin}} w^{3/2} ,
\ee
where we have also used (\ref{wpin}). 
This result can be cast in a more transparent form if we express $w$ in terms of the magnetic field
perturbation $\delta B$. Using equation~(\ref{wA1}) from Section~\ref{sec:multifluids} we have
\be
\tau_{\rm mf} \sim \frac{1}{8} x_\rp P  \frac{v_\Lambda^{1/2} v_\rA^{3/2} 
}{w^2_{\rm pin}} \left ( \frac{\delta B}{B} \right )^{3/2} .
\ee
The key unknown in this expression is the perturbed magnetic field $\delta B$. Given that fluxtube cutting 
operates for a mode amplitude above the threshold required by vortex unpinning it makes sense to normalise $\delta B$ 
to its value $\delta B_{\rm pin}$, see eqn.~(\ref{dBpin}) (obviously in the fluxtube cutting regime $\delta B > \delta B_{\rm pin}$). 

We can then obtain our ``final'' result for the mutual friction timescale,
\bear
&& \tau_{\rm mf} \sim  \frac{1}{8} x_\rp P \left ( \frac{v_\Lambda}{w_{\rm pin}} \right )^{1/2} 
\left ( \frac{\delta B}{\delta B_{\rm pin}} \right )^{3/2} ,
\label{taumf1}
\\
\nonumber \\
&& \qquad \approx 3\, x_5  \rho_{14}^{-1/2} \left ( \frac{P}{10\,\mbox{s}} \right )  \, \left ( \frac{\delta B}{\delta B_{\rm pin}} \right )^{3/2}
 B_{15}^{-1/4} \, \mbox{s} .
 \label{taumf2}
\eear
This result implies that a global magnetar oscillation will experience strong damping once the amplitude
exceeds the one required for vortex unpinning ($\delta B > \delta B_{\rm pin}$). 
The cutting mechanism shuts down for any $\delta B < \delta B_{\rm pin}$.



\subsection{Internal damping: vortex-electron friction}
\label{sec:mf}

There is a second type of mutual friction operating in our two-fluid system, arising from the dissipative scattering of
electrons by the magnetized neutron vortices in the core. In fact, this is the mutual friction mechanism usually considered
in the superfluid neutron star literature, see~\citet{als88,NA06}. It corresponds to a force exerted on unit vortex length 
\be
\mathbf{f}_{\rm D} =  \rho_\rn \kappa \cR ( \bv_\rp -  \bu_\rn) 
\ee
when the vortex moves with respect to the electrons. Note that the drag coefficient $\cR$ appearing in this expression is constant.  

Assuming that this is the only frictional force exerted on the vortex, we can arrive at the mutual friction
force~(for details see~\citet{HV,NA06}),
\be
\bF_{\rm mf} =  -2\Omega_\rn \left [  \cB  \left \{ \mathbf{\hat{z}} \times (\mathbf{\hat{z}} \times \bw  ) \right \} 
+ \cB^\prime (\mathbf{\hat{z}} \times \bw  ) \right ] ,
\ee
where $\cB = \cR/(1+\cR^2)$ and $\cB^\prime = \cR^2/(1+ \cR^2)$.  The detailed calculation of $\cR$~\citep{als88,NA06} suggests that $\cR \approx 4\times 10^{-4} $
and therefore 
\be
\cB \approx \cR \ll1 , \qquad \cB^\prime \approx \cB^2 \ll \cB .
\ee
 
We can estimate the damping timescale of Alfv\'en modes as a result of electron mutual friction by following the same logic as in the preceding section.   
The damping rate is (we note that the $\cB^\prime$-piece of $\bF_{\rm mf}$ is not dissipative)
\be
\dot{E}_{\rm mf} \sim 2\Omega \rho \cB \int_0^R dr r^2 w^2 .
\ee  
This leads to the damping timescale
\be
\tau_{\rm mf} \sim \frac{1}{2} \frac{x_\rp P}{\cB} \approx
630 \, x_5 \left ( \frac{P}{10\,\mbox{s}} \right )  \left ( \frac{4\times 10^{-4}}{\cB} \right ) \,\mbox{s} .
\label{taumf_e}
\ee
This is significantly longer than the timescale associated with fluxtube cutting, but short enough to be relevant for the evolution of
magnetar QPOs. 

As we have pointed out, the derivation of (\ref{taumf_e}) assumes that no other frictional force acts on individual
vortices. The vortices in the real system are likely to experience a mutual friction that is a combination of electron friction and
of the force due to fluxtubes. Both mechanisms operate when the vortices can move with respect to the fluxtubes (which are
frozen in the proton-electron fluid) and their relative strength would depend on the specific geometry between the 
vortex and fluxtube arrays (see for example~\citet{SA09}). In that case the mutual friction damping should be given by the
shortest timescale between (\ref{taumf2}) and (\ref{taumf_e}).


\subsection{Internal damping: other mechanisms}

The stellar interior allows for additional dissipation channels -- the most conventional ones are standard shear and bulk viscosity (these
are represented by the viscous force $\bF_{\rm visc}$ in the Euler equations of Section~\ref{sec:multifluids}). 
However, it is easy to see that in the temperature regime where the observed magnetars are expected to reside (i.e. core temperature 
$T \sim 10^8\,\mbox{K} $, see for instance~\citet{HGA12}) none of these mechanisms is of any importance. To see this, we can use approximate, back-of-the-envelope, 
viscous timescales derived from $F_{\rm visc}$ as in, for example, \citet{CL87}. These are
\be
\tau_{\rm sv} \sim \frac{\rho L^2}{\eta}, \qquad \tau_{\rm bv} \sim \frac{\rho L^2}{\zeta} ,
\ee   
where $\eta, \zeta$ are the coefficients of shear and bulk viscosity, respectively. For the former parameter we use the value appropriate for
superfluid matter, associated with electron collisions~\citep{NA05} $\eta =  1.6\times 10^{19}\, x_5^{3/2} \rho_{14}^{3/2} T_8^{-2}  $.
For the bulk viscosity coefficient we use Sawyer's standard normal-matter formula~\citep{sawyer}.
(This overestimates the damping rate because it does not account for the suppression of the beta-equilibrium chemical reactions 
due to neutron and proton superfluidity). 
 
We then obtain 
\be
\tau_{\rm sv} \sim 6 \times 10^6\, \rho_{14}^{-1/2} x_5^{-3/2} L_6^2 T_8^2 \, \mbox{s} .
\ee
The bulk viscosity timescale is even longer so we do not write it explicitly. These results clearly show that neither mechanism is a factor in the evolution
of magnetar oscillations. 

There is a third possible damping mechanism, most commonly known in the context of the $r$-mode instability~\citep{NAKK01}: the viscous boundary
layer formed at the base of the crust due to the discontinuity in the mode's velocity field. This mechanism is rather important for $r$-modes, severely 
limiting their `instability window' in the temperature regime where neither shear nor bulk viscosity is efficient.  However, we can show that
the boundary layer poses no threat to magnetar oscillations simply because it is not formed. 

This can be deduced by studying a simple plane-parallel model of the crust-core interface (as in~\citet{BU00})) or by using the results of
~\citet{mendell01}. It is then found that the magnetic field can seriously modify the physics of the layer and, among other things, 
make the layer thicker (in the radial direction). For this to happen the magnetic field must satisfy $v^2_\rA / \omega\nu \gg 1$
where $\nu = \eta/\rho_\rp$ and $v_\rA$ is the superconducting Alfv\'en speed (\ref{vA2}).  We find 
\be
\frac{v^2_\rA}{\nu \omega} \sim 10^{8}\, x_5^{-5/2} \rho_{14}^{-3/2} H_{15} B_{15} T_8^2   \left (\frac{100\,\mbox{Hz}}{f_{\rm mode}} \right ) .
\ee
Clearly, the boundary layer is dominated by the magnetic field. Its thickness is given by~\citep{mendell01}
\be
\delta_{\rm VBL} \approx \frac{2 v^3_\rA}{\omega^2 \nu} .
\ee
We should obviously have $\delta_{\rm VBL} \ll R$ for the notion of the layer to make sense. Inserting numbers in this formula,
\be
\delta_{\rm VBL} \sim 10^{14}\, x_5^{-3} ( H_{15} B_{15} )^{3/2} T_8^2 \rho_{14}^{-2}  \left (\frac{100\,\mbox{Hz}}{f_{\rm mode}} \right )^2   
\, \,~ \mbox{cm} .
\ee
This length is much bigger than the stellar radius throughout the relevant parameter space. 
We therefore conclude that a boundary layer does \emph{not} form during magnetar oscillations.


\subsection{Gravitational waves: emission and damping} 
\label{sect:GWs}

Gravitational wave emission from the excited modes is of interest for two reasons.  Firstly,  the gravitational waves will themselves damp the oscillations, so we need also to estimate the decay timescale for gravitational radiation damping. Secondly, the gravitational wave emission may be detectable, and so could provide independent measurements of frequencies and damping times.  

Gravitational wave measurements of the damping times would be particularly useful, as the lifetimes of the QPOs as observed via X-rays is probably \emph{not} a good indicator of the lifetime of the stellar oscillation itself.  This is made clear by the fact that certain QPOs appear briefly only in the \emph{middle} of the X-ray burst, whereas the modes themselves are presumably excited once, at the beginning of the event.  Also, the X-ray QPOs are visible as the modulations of the light curve produced by the relativistic fireball created by the initial event, so that once the fireball has faded and the X-ray flux fallen, the X-ray modulations will no longer be detectable, even if the star itself continues to oscillate.  The lifetime of the X-ray QPOs should therefore be taken as \emph{lower limits} on the lifetimes of the modes themselves.  A `direct' gravitational wave observation of the modes would suffer no such selection effect, and give a robust measure of the mode decay timescale. 

We will begin by estimating the gravitational wave damping timescale.   The modes we consider here are all of low frequency, in the sense of having frequencies well below the dynamical fundamental $f$-mode frequency of $\sim 1-2$ kHz.  As such, they induce very small density perturbations, consistent with the neglect of such perturbations in Section \ref{sec:dynamics}, and to this level of approximation will not radiate mass quadrupole radiation.  However, the modes have significant non-zero velocity perturbations, and so there will be mass current quadrupole radiation, the luminosity of which  we can easily estimate.

Using the formalism of \cite{thor80}, the energy flux for mass current quadrupole radiation is (to order-of-magnitude precision) $\dot E \sim  (\omega^3 S)^2$
where $S$ denotes the mass current quadrupole, which can be approximated by a simple volume integral over the mass current  $\rho{\bf v}$ over the star:
\be
S \sim \int |{\bf r} \times (\rho {\bf v}) r| \, dV .
\ee
These will be contributions from the both the proton mass current $\rho_\rp {\bf v_\rp}$ and the neutron mass current $\rho_\rn \bf v_\rn$.

In the case of Alfv\'en modes, it is easy to show using the results of Section \ref{sec:multifluids} that there is a partial cancellation between the two mass currents, with a net current of $\sim \rho_\rp {\bf v_\rp} m_\rp^*/m_\rp$.  We can insert this into the equation above for $S$ above and then computing the gravitational wave luminosity.  Combing with the mode energy of equation  (\ref{Emode2}) leads to a damping time
\be
\tau_{\rm GW} \sim  \frac{1}{x_\rp M R^4 \omega^6} 
\sim 10^{7} x_5^{-1} M_{1.4}^{-1} R_6^{-4} \left (\frac{f_{\rm mode}}{100\,\mbox{Hz}} \right )^{-6}\,  {\rm \, s} .
\ee
The corresponding timescale for the damping of crustal oscillations is very similar.  Comparing with the various internal and external damping timescales calculated previously, we see that gravitatinal radiation is \emph{not} a significant source of damping for our magnetar QPOs.  

The detectability of gravitational waves from magnetar flares has already been considered in the literature.  Some optimistic upper limits were obtained by looking at how much energy could be channelled into modes (see e.g. \citet{CO11}).  Probably more realistically, detailed numerical simulations have also been performed, see e.g. \citet{ZLK12}, who did indeed find excitation of low-frequency modes.   A key uncertainty was the damping times of these modes, something which the (short duration) numerical simulations were unable to compute.  Scaling the results of Figure 3 of \citet{ZLK12} to a damping time of, say, $100$ s, typical for some of the damping mechanisms considered above, we see that detection by a third generation detector is possible, and detection by a second generation (advanced) detector cannot be ruled out, although a large magnetic field strength $\sim 10^{16}$ G  would be required.

\section{Concluding discussion}
\label{sec:conclusions}

Our three `evolutionary pathways' make several predictions with a bearing on the observational signature of magnetar QPOs.   These were discussed in Section~\ref{sec:paths} and can be summarized as follows: 

(i) The damping timescale due to the combined action of superfluid vortex mutual friction and magnetospheric Alfv\'en flux can move both \textit{up} and \textit{down} during the evolution.  The \emph{observed} lifetimes of X-ray QPOs  range from tens to hundreds of seconds  \citep{SW06, HNK11}.  Our calculated timescales can accommodate such a range, although we emphasise that the observed X-ray lifetimes may not accurately reflect the lifetime of the stellar oscillation  itself.

(ii) The frequency of the Alfv\'en spectrum can only be shifted \textit{upwards} to higher frequencies during the evolution as a result of the restoration of neutron superfluidity in the stellar core. 

Clearly, our work suggests that analysis of the observational data should allow for the possibility of both frequency and damping timescale variations for the QPOs.  On the theoretical side, our work indicated that a realistic ``magnetar asteroseismology" programme should involve many more elements than the mere identification of the observed frequencies with a `fixed' theoretical oscillation spectrum.

Our model can be (and should be) improved in many ways. An obvious missing ingredient is the connection between the
oscillation amplitude in the star and that of the emitting region outside the star (the ``fireball"). Apart from the obvious need of improving the existing 
magnetar oscillation models, special attention must be given to the damping mechanisms discussed here, i.e. vortex mutual friction and electromagnetic
losses into the magnetosphere. Non-linear mode couplings (not considered here) could be an important part of the story too, especially in the early 
post-flare stages when the mode amplitudes are presumably largest.   Another key unknown is that of the initial conditions: which modes are excited and what are their relative amplitudes? Answering these kinds of
questions goes hand in hand with understanding the detailed nature of the flare's trigger-mechanism.
 
On a more technical level, while we were able to confirm the reliability of our simple estimates for the magnetospheric damping timescales of Sections \ref{sec:alfven} and \ref{sect:md:cm}, using full surface integrals for the flux, and volume integrals for the mode energy, our estimates for mutual friction damping timescales of Sections \ref{sec:cutting} and \ref{sec:mf} were done only in a simpler order-of-magnitude fashion.  More detailed calculations, using the full mode eigenfunctions, are needed to test the reliability of our estimates.
 
Another interesting and novel problem that needs attention concerns computing mode frequencies and damping rates when there are sharp transitions in time and space between superfluid/non-superfluid phases, and between pinned/non-pinned phases.  We have provided some simple toy model estimates of how these might tend to smooth out the transitions between the three regimes of interest to us, but full mode calculations that  incorporate such transitions are clearly needed to properly understand this issue.
 
It is clear that the topical area of magnetar flares is a very challenging one but also one that offers a unique insight into these extreme objects, potentially offering information on both the processes internal to the star, and the star's interaction with its magnetosphere.


\section*{Acknowledgements}

It is a pleasure to thank Lars Samuelsson for useful discussions. 
KG is supported by the Ram\'{o}n y Cajal Programme of the Spanish Ministerio de Ciencia e Innovaci\'{o}n 
and by the German Science Foundation (DFG) via SFB/TR7. 
DIJ acknowledges support from STFC via grant number ST/H002359/1, and both authors acknowledge travel support from CompStar
(an ESF-funded Research Networking Programme).


\end{document}